\documentclass[a4paper,oneside,11pt]{article}
\usepackage{graphicx}
\usepackage{color}
\usepackage{amsmath, amssymb}
\usepackage{mathabx}
\usepackage{amsthm}
\usepackage{float}
\usepackage{mathrsfs}
\usepackage[font=scriptsize,labelfont=bf]{caption}
\definecolor{verde hierba}{rgb}{0,1,0}

\providecommand{\keywords}[1]
{\small\textbf{\textit{Keywords---}} #1}

\title{\textbf{\emph{On the Concept of Frequency 
in Signal Processing}}}

\author{\small{Mois\'es Soto-Bajo, 
Andr\'es Fraguela Collar, 
Javier Herrera Vega, 
Ra\'ul Felipe Sosa}}
\date{\small{\today}}

\begin{document}

\maketitle

\begin{abstract}
Frequency is a central concept 
in Mathematics, Physics, and Signal Processing. 
It is the main tool for describing 
the oscillatory behavior of signals, 
which is usually argued to be 
the manifestation of some of their key features, 
depending on their nature. 
For instance, 
this is the case of Electroencephalographic signals. 
Hence, 
frequency is substantially present 
in the most common methodologies for analyzing signals, 
as the Fourier Analysis or the Time-Frequency Analysis. 

However, 
in spite of its importance 
as a keystone in Signal Processing, 
and its seemingly simple meaning, 
its mathematical foundation is not 
as straightforward as it may seem at first glance. 
A naive interpretation 
of the different mathematical concepts modelling frequency 
can be misleading, 
as their actual meanings 
essentially differ 
from the intuitive notion 
which are supposed to represent. 

In our opinion, 
this circumstance should be taken into account 
in order to develop 
appropriate signal analyzing and processing tools 
in some applications. 
In the current text 
we discuss this topic, 
with the main goal 
to draw the attention 
of the mathematical and engineering community 
to this point, 
often overlooked. 
\end{abstract}


\bigskip
\hrule





\begin{center}
\textbf{Motivation}
\end{center}

The concept of ``frequency'' is almost ubiquitous 
in Signal Processing and the many disciplines 
which make extensive use of it. 
Among them, 
the Bioengineering, 
in which frame the processing and analysis 
of bioelectric signals is a main task. 
Particularly, 
in Neurology and Cardiology, 
respectively, 
the appropriate analysis of 
Electroencephalographic (EEG) 
and Electrocardiographic (ECG) signals 
is essential in order to understand 
the brain and heart electrophysiology 
(\cite{SL:05}, \cite{NS:06}, \cite{SchLdS:12}, 
\cite{SCh:13}, \cite{CAMcS:06}, \cite{To:93}). 

Specially interesting is the case of Neurophysiology. 
In spite of there is not yet total consensus 
in the clinicians and researchers scientific community 
about the specific signal patterns 
associated to healthy states or each illness, 
and how they can be trustfully used 
in Clinical Neurophysiology, 
there is ample evidence that 
this kind of biometric signals 
play an important role 
(and will play an increased one) 
in assessing the physiological function 
in the brain 
(\cite{Hir:13}, \cite{Bab:20}). 

EEG signals in a healthy subject, 
apart from noise and artifacts, 
and many other considerations, 
are supposed to be the 
superposition or juxtaposition 
of brain rhythms 
(\cite{SL:05}, \cite{SCh:13}). 
From the point of view of Neuroscience, 
brain rhythms are the manifestation of 
the collective and synchronous activity 
of millions of neurons, 
and depend on many factors. 
However, 
they are characterized by 
frequency and amplitude ranges 
(\cite{SL:05}). 
Clinical experts are trained on visual inspection of EEG, 
mainly based on examining 
the oscillations of the signal, 
frequently by counting the zero crossings 
(\cite{NS:06}, \cite{SCh:13}, \cite{SchLdS:12}). 
Simplifying, 
oscillations are outlined by 
the ups and downs of the signal, 
where duration and height between them 
determine the frequency and amplitude, 
respectively. 

In the absence of a rigorous definition 
(in a mathematical sense), 
ignoring the neurophysiological aspects and 
from a mathematical or signal processing point of view, 
brain rhythms are oscillatory signals 
which possess a close-to-uniform oscillation rate, 
which fluctuates inside a specific range, 
called frequency band 
(typically known as 
$\delta,\theta,\alpha,\beta,\gamma$ 
rhythms). 
Moreover, 
there is not even yet complete consensus 
about the limits of the frequency bands 
(\cite{Bab:20}). 

Consequently, 
from the perspective of Signal Processing, 
the basic goal is to decompose signals 
into rhythms, 
or at least being able to analyze 
these ``components'' or 
differently ``oscillatory-behaved'' manifestations. 
Because of the lack of 
firmly established and consensual definitions, 
this turns to be a hard task, 
far from being completely understood. 
In view of the basic (or ``naive'') description of rhythms, 
it is not surprising that 
frequency-based analyses 
are in the heart of this kind of analysis, 
and many techniques and procedures 
giving place to automatized analytical algorithms 
have been developed to process EEG signals 
(\cite{SL:05}, \cite{NS:06}, \cite{SchLdS:12}, 
\cite{SCh:13}, \cite{To:93}, \cite{LBBM:12}). 

One question arises: 
how these fuzzy notions 
could be mathematically modeled? 
Well thought-out, 
this is a bewildering and challenging question. 
The customary answer follows the following paradigm: 
the current signal is a relation 
between the independent variable, 
let say time, 
and the main magnitude 
(in the EEG case, 
voltage difference). 
But there is another, 
the spectral representation of the signal 
(depending on the chosen technique), 
which is another function 
which relates to each frequency 
the amplitude corresponding 
to the basic oscillation 
at this frequency; 
that is, 
the strength at which 
this harmonic participates or takes part 
in the composition of the signal. 
Clearly, 
the first description is elementary; 
it is completely linked to the perception or measurement. 
However, 
the second one is less intuitive, 
and depends strongly on 
the definition of frequency itself 
and the analysis procedure performed. 

According to this description, 
rhythms need to be understood 
in terms of the chosen spectral representation 
and the corresponding concept of frequency. 
However, 
it is clear that 
any performed analysis should respect 
the neuroscientists paradigm, 
in terms of oscillations, 
and the subsequent components 
in which signals are split 
should be meaningful for the specialists. 
Namely, 
the used concept of frequency, 
in every case, 
should not deviate a lot 
from this perspective. 

Having said that, 
it is clear that 
the used concept of frequency is critical 
in this framework. 
But, 
what do we exactly mean/understand by ``frequency'' 
in Signal Processing? 
How is it related to oscillations? 
Is it successful or adequate 
for answering the key questions 
in Neuroscience, 
or other disciplines? 


\begin{center}
\textbf{Classical Fourier analysis}
\end{center}

Undoubtedly, 
Signal Processing borrowed the concept of frequency 
from Physics and Mathematical Analysis, 
specifically from Fourier Analysis. 
The sinusoidal basic functions 
are given by the relations 
$t\longmapsto\sin(2\pi\omega t)$ 
or $\cos(2\pi\omega t)$, 
or in its complex form, 
$t\longmapsto e^{2\pi i\omega t}$. 
Here $i$ is the imaginary unit, 
and the real numbers 
$t,\omega\in\mathbb{R}$ 
represent time (in seconds) 
and frequency (in hertz), 
respectively. 
The physical interpretation is 
the angular frequency or angular speed 
(but measured in hertz) 
in the corresponding rotary motion, 
and gives the rate of change of the phase, 
measuring the angular displacement 
in cycles per unit time. 
Thinking in terms of oscillations, 
frequency measures the speed 
of this oscillatory behaviour. 
The larger $|\omega|$ is, 
the more cycles are traveled 
per unit of time. 
Actually, 
$|\omega|$ is the number of cycles 
performed per unit of time itself. 
Note that this interpretation 
(in a strict sense) 
is naturally restricted to pure sinusoidal signals. 

In Harmonic Analysis, 
trigonometric polynomials 
(linear combinations of sinusoidal functions) 
are studied, 
and more general functions 
are sought to be represented and developed 
somehow in terms of basic sinusoidal functions. 
These representatives are defined by 
the set of frequencies involved, 
and the corresponding amplitudes 
accompanying the basic sinusoidal components, 
in the form of sequences of coefficients or functions, 
depending on the nature of the frequency set 
(discrete or continuous). 
Apart from the selected set of frequencies, 
these amplitudes, 
which are the response in frequency of the signal, 
or spectral representation of the function, 
define the representation 
and are sought to characterize the signal. 

Hence, 
the representation process is split into two ones: 
the analysis, 
which consists in obtaining the representative 
given by the corresponding amplitudes, 
and the synthesis, 
which consists in recovering the signal 
from its representative, 
both performed by suitable operations. 
These representations usually take the form of 
discrete or continuous expansions, 
depending on the nature of the frequency set, 
and series or integrals 
respectively appear in the definition 
of such representations. 

Basically, 
classical Fourier Analysis consists of 
two main areas: 
the Fourier Series, 
and the Fourier Transform 
(we refer to \cite{Z:59}, 
\cite{Ka:68}, 
\cite{DMcK:72} for details). 

The Fourier Series theory 
deals with periodic functions (signals), 
and provide an effective way 
for analyzing and synthesizing them. 
Here, 
the most remarkable result is 
the Plancherel theorem: 
the suitably normalized trigonometric system 
$\{e^{2\pi ik\cdot/(b-a)}/\sqrt{b-a}\,:\,k\in\mathbb{Z}\}$ 
is an orthonormal basis 
of the Lebesgue space of 
locally square summable $(b-a)$-periodic functions, 
where $\mathbb{Z}$ is the set of integers. 
Consequently, 
any such periodic function 
$f$ 
writes as the corresponding Fourier series, 
where the information of $f$ 
is encoded into its Fourier coefficients 
which, 
due to the orthonormality, 
are given by the following expression, 
$$
\widehat{f}(k)=
\int_{a}^{b}f(t)\frac{e^{-2\pi ikt/(b-a)}}{\sqrt{b-a}}\,dt
\qquad(k\in\mathbb{Z})\,,
$$
and satisfy the Parseval identity: 
$$
\int_{a}^{b}|f(t)|^{2}\,dt=
\sum_{k\in\mathbb{Z}}|\widehat{f}(k)|^{2}\,.
$$

Analogously, 
the Fourier Transform deals with 
functions defined on 
the whole real line $\mathbb{R}$. 
For an integrable function $f$, 
the Fourier transform is given by: 
$$
\widehat{f}(\xi)=
\int_{-\infty}^{\infty}f(t)e^{-2\pi i\xi t}\,dt
\quad(\xi\in\mathbb{R})\,,
$$
It turns to be a unitary operator 
defined in the Lebesgue space of square summable functions 
$L^{2}(\mathbb{R})$, 
satisfying the Plancherel identity, 
$$
\int_{-\infty}^{\infty}|f(t)|^{2}\,dt=
\int_{-\infty}^{\infty}|\widehat{f}(\xi)|^{2}\,d\xi\,.
$$

Note that in the first case, 
the set of frequencies is $\mathbb{Z}$, 
we represent periodic functions, 
the analysis process consists on 
computing the sequence of Fourier coefficients, 
and the synthesis process consists on 
summing the Fourier series of the function. 
In the second case, 
the set of frequencies is $\mathbb{R}$, 
we represent functions in $L^{2}(\mathbb{R})$, 
and the analysis and synthesis processes 
are performed by the Fourier and inverse Fourier transform, 
respectively. 

These objects extend the concept of frequency 
from pure sinusoidal functions 
to more general signals. 
Now, 
frequency is a parameter 
which indexes the set of 
responses of the signal 
when it is faced against 
a pure oscillation, 
resulting the corresponding value 
of the amplitude in the representation: 
the Fourier coefficient $\widehat{f}(k)$ 
or the Fourier transform value $\widehat{f}(\xi)$, 
respectively. 

But there is (always) a price. 
In this framework, 
the exact meaning of the concept ``frequency'' 
is hidden behind the frequency variables 
$k$ and $\xi$, 
respectively. 
It is no longer an 
oscillation speedometer 
(although it is supposed to be). 
This is the point we want to highlight here, 
which plunges into the deep nature 
of the greatest problem in time-frequency analyses, 
since it is inherent 
to the central concept of frequency itself. 

It is easy to see 
some eloquent symptoms of this: 
one of the main tasks in signal processing 
is time-frequency localization, 
or identification of 
the characteristic oscillations of signals, 
and where they occur. 
First of all, 
it is worth noting that, 
in real applications, 
one (always) deals with 
discrete (actually finite) signals 
on a finite interval. 
Nevertheless, 
the corresponding theory for analyzing finite signals 
runs parallel to the continuous one. 

The analysis is mostly performed by 
(or based on) 
linear functionals. 
Under orthogonality conditions 
(or similar, 
as in Frame Theory 
-for the theory of frames 
and Riesz bases 
we refer to \cite{Ch:03}-), 
one computes the inner product 
of basic harmonics 
on a finite window. 
Thus, 
for any $\xi,\eta\in\mathbb{R}$ 
one has 
$$
\int_{a}^{b}e^{2\pi i\eta t}e^{-2\pi i\xi t}dt=
(b-a)\,e^{\pi i(\eta-\xi)(a+b)}\,
\frac{\sin(\pi(\eta-\xi)(b-a))}{\pi(\eta-\xi)(b-a)}\,.
$$
Note that this can be interpreted 
as the Fourier transform 
of a windowed pure tone, 
also. 
This implies that, 
computed in this way, 
basic tones interfere each other 
in the whole spectrum of frequencies, 
excluding a discrete sequence of pairings 
in harmonic consonance: 
$(\eta-\xi)(b-a)\in\mathbb{Z}$. 
Also, 
this means that 
this Fourier transform 
is entirely spread over $\mathbb{R}$, 
not condensed at the corresponding frequency $\eta$. 
In some sense, 
this interference disappears 
when one considers the ``whole signals'', 
supported in the whole real line, 
but this occurs 
outside the frontiers of $L^{2}(\mathbb{R})$ 
because of the failing of integrability. 
This results very inappropriate since, 
apart from mathematical technicalities, 
recall we are usually interested in 
studying signals on finite intervals. 
But it turns out that 
Fourier transform is designed 
to handle signals 
defined in $\mathbb{R}$. 
You can restrict the support 
of the signal, 
but you don't fool Fourier transform. 

The other side of the same coin 
are the Fourier coefficients on an interval, 
where orthogonality relations 
are obtained at the price of 
neglecting the dissonant frequencies. 
As a consequence, 
Fourier series scatters 
the energy of a dissonant pure tone 
through the whole spectrum, 
due to the Parseval identity, 
giving place to a 
complete spectrum of spurious amplitudes, 
with no physical meaning at all 
(see the equation above 
with $\xi,\eta$ not satisfying 
$(\eta-\xi)(b-a)\in\mathbb{Z}$). 
And discordant frequencies 
are not superfluous at all; 
they can occur naturally, 
as the superposition of 
independent, 
asynchronous and uncoordinated, 
but concurrent, 
contributor oscillations. 
In this case, 
the frequency analysis 
definitely perverts 
the physical nature 
of the signal. 
It is mathematically correct, 
but not informative 
(or worst, 
misinformative). 

Another parallel ideas, 
shaped like uncertainty principles, 
are well known 
(see \cite{dB:67}, 
\cite{Co:95}, 
\cite{Fo:97}, 
\cite{Gro:01}, 
\cite{DS:89}, 
\cite{Sm:90}, 
\cite{FT:20}). 
In the framework of 
Fourier transform and series, 
which were designed/developed for other, 
different in nature, 
purposes, 
these drawbacks cannot be avoided. 
Therefore, 
these and other 
Fourier Analysis-based tools 
for time-frequency localization, 
which constitute the main corpus 
in this discipline, 
suffer of this same illness. 


\begin{center}
\textbf{Beyond Harmonic Analysis: 
Time-Frequency Analysis}
\end{center}

Harmonic (or Fourier) analysis 
is a classical discipline, 
which has turned to be the cornerstone 
of many others, 
in pure and applied sciences. 
Concerning to time-frequency localization, 
some techniques have been developed 
based on these concepts, 
as the family of Fast Fourier Transforms (FFT) algorithms 
(see \cite{CT:65}, 
\cite{CLW:67}, 
\cite{HJB:85}, 
\cite{Coo:87}), 
or the Theory of band-limited functions 
and the Prolate Spheroidal Wave Functions (PSWF) 
(see \cite{SP:61}, 
\cite{LP:61}, 
\cite{LP:62}, 
\cite{Sl:64}, 
\cite{ORX:13}). 
These tools have been successfully applied 
in many applications. 
However, 
in spite of their remarkable properties, 
there are also disadvantages 
in their use. 
For instance, 
PSWF are computationally challenging 
(\cite{WS:05}). 
Anyway, 
we want to focus here on the fact 
that all these techniques 
depend on the concepts of frequency 
just explained above, 
and consequently share their faults. 

Going further, 
beyond the classical harmonic framework, 
much effort has been devoted to developing 
alternative tools and procedures 
for capturing the main features of a signal
by simultaneously localize it 
in time and frequency. 
In other words, 
the point is to being able to identify 
the pairs of time windows and frequency bands 
in which characteristic oscillations of the signal occur. 
These different approaches can be included 
under the common name of Time-Frequency Analysis 
(see \cite{Co:95}, \cite{Gro:01}, \cite{Boa:15}). 
In contrast to classical Harmonic Analysis, 
Time-Frequency Analysis treats 
(or attempts to) 
the time and frequency variables equally, 
as primary concepts, 
and works with them simultaneously. 

Many techniques have been developed 
with this flavor: 
the Short-Time Fourier Transform (STFT), 
in which a suitable 
(very localized in time and frequency) 
window function 
is used to ``redefine'' the Fourier Transform; 
the Spectrogram, 
the Ambiguity function, 
the Wigner-Ville distribution, 
and other Quadratic time-frequency representations, 
usually closely related to the STFT, 
which can be thought as 
energy densities in the time-frequency plane, 
which are not completely satisfactory 
(see \cite{Gro:01}); 
and also the Gabor frames, 
where discrete expansions 
in terms of time-frequency shifts on a lattice 
of a suitable window function are considered. 
All these tools have been 
successfully applied to many processing tasks, 
as feature extraction, 
separation of signal components, 
and signal compression. 
See \cite{Co:95}, \cite{Gro:01}, \cite{Boa:15} 
for more details. 

In some sense, 
the Time-Frequency techniques 
define a local frequency spectrum 
which is supposed to report 
the strength of each oscillation 
in the signal. 
However, 
in spite of their success, 
they borrow their concept of ``frequency'' 
from Fourier analysis. 
That is, 
the response in frequency 
is obtained by facing up 
(by integration) 
an averaged portion of the signal 
(by means of window functions) 
to the exponential kernel, 
which encodes the oscillation speed; 
that is, 
the chosen frequency. 
Consequently, 
the deficiencies of this approach 
are inherited from 
the corresponding Fourier transform ones; 
they cannot break out of the limits 
that classical Fourier Analysis imposes. 
It is worth noting that 
the frequency variable 
is not actually an absolute primary concept; 
the way it appears in each transform 
(and only that) 
accurately determines its precise meaning, 
not the name with which it was coined. 

Some symptoms of this situation 
are well known actually 
(although the diagnosis maybe not so well). 
Two main principles are ubiquitous 
in Time-Frequency Analysis 
(they refer to the 
time and spectral representations of the signal): 
1. The smoothness-decay duality: 
if one representation is smooth, 
the other one decays. 
2. The uncertainty principles: 
the two representations cannot be 
simultaneously well localized. 

About the first one, 
at first sight 
it seems reasonable that 
sudden jumps in the signal 
produce respective manifestations 
at high frequencies. 
But consider the following example. 
Let the square signal 
given by the sign of $\sin(2\pi t)$. 
How would you describe 
its periodicity, 
its oscillation, 
in terms of frequency? 
How do you think 
a neurophysiologist would do it? 
Now compute its Fourier transform 
and compare. 

Concerning the uncertainty principles, 
they constitute a well studied topic 
(see \cite{dB:67}, 
\cite{Co:95}, 
\cite{Fo:97}, 
\cite{Gro:01}, 
\cite{DS:89}, 
\cite{Sm:90}, 
\cite{FT:20}): 
the Heisenberg-Pauli-Weyl inequality, 
the Donoho-Stark uncertainty principle 
(about the concentration of a signal 
and its Fourier Transform, 
in a wider sense: 
on the real line, 
periodic signals, 
discrete signals, 
on LCA groups. 
See \cite{Co:95}, 
\cite{Gro:01}, 
\cite{DS:89}, 
\cite{Sm:90}); 
Lieb's inequalities 
(for the STFT); 
the radar uncertainty principle 
(for the Ambiguity function); 
the Wigner-Ville distribution uncertainty principles, 
or the results on the positivity 
of smoothed Wigner-Ville distributions; 
the Balian-Low theorem 
and density theorems for Gabor frames, 
etc. 

About their meaning, 
as far as we are concerned, 
the uncertainty principles 
make the existence of an absolute and ideal 
concept of instantaneous frequency 
(or spectrum) 
impossible 
(see \cite{Gro:01} for a detailed discussion), 
in the following sense: 
there is no (square-summable) function 
concentrated on an arbitrarily small interval 
with Fourier transform 
also concentrated on an arbitrarily small band. 
However, 
in this regard, 
it is extremely interesting to recover 
the discussion in \cite{Sl:76} 
(see also \cite{Sl:83}) 
about the hypothesis of bandlimitedness 
and the mathematical modelling 
of ``real world'' signals, 
and the $2WT$-theorem. 

Nevertheless, 
there are exceptions. 
Remarkable ones are the Wilson basis 
and the Malvar basis 
(see \cite{DGM:86}, 
\cite{DJJ:91}, 
\cite{CM:91}), 
or more generally 
the local Fourier bases. 
Being extremely simplistic, 
the basic idea is 
to consider simultaneously 
symmetric frequencies 
($\xi$ and $-\xi$), 
replacing complex exponentials 
by sine and cosine functions, 
and overlap block families 
with suitable window functions 
(see 
\cite{ABF:94}, 
\cite{Gro:01}, 
\cite{HW:96}). 
This kind of bases 
escape from the uncertainty principle. 
But again, 
due to its structure, 
as orthogonal bases 
of windowed Fourier series, 
suffer of the same deffects 
as the Fourier series. 
Furthermore, 
the window smooths 
the edge effects, 
but also totally reshapes 
the sinusoidal waveforms, 
intimately related to the notion of oscillation. 

Another case of interest is 
Wavelet theory. 
Wavelets have been applied with great success 
in Signal processing 
and many other disciplines. 
It is a fact that 
there are well-localized in time-frequency wavelets, 
very suitable for time-frequency analysis 
(see \cite{Da:92}, 
\cite{HW:96}, 
\cite{WS:04}). 
Nevertheless, 
not all that glitters is gold. 
In this case, 
in spite of Fourier Analysis 
is a fundamental tool in Wavelet theory 
(the fact that Fourier transform interchange 
translations and modulations 
basically explains the interplay 
between Wavelet and Gabor Analyses), 
actually the concept of frequency 
is in some way replaced 
by the concept of scale (or resolution), 
as modulations are replaced by dilations, 
in this theory. 
The scale spectrum grows geometrically 
(at least for the Discrete Wavelet Transform, 
the more relevant in practice), 
as powers of two in the classical setting. 
Also, 
the mesh refines according to scale. 
These circumstances induce that 
``dissonant'' features 
(that is, 
those whose scale and position 
fall between the lattice ones) 
will be scattered through 
several spurious components, 
according to energy-preservation laws. 
On the other hand, 
there is relatively much freedom 
in choosing the waveform 
for wavelets, 
which do not need at all 
to be sinusoidal 
(actually, 
they do not are waves, 
but ``little waves'', 
in virtue of the so called 
oscillation property 
$\int\psi=0$). 

And last but not least, 
we should mention 
the Instantaneous Frequency. 
This concept has a long history, 
but its use and 
indeed its own nature 
have been quite controversial, 
in spite of it has been successfully applied 
in many situations, 
specially for the analysis 
of nonstationary signals. 
Instantaneous frequency is a magnitude 
which is supposed to measure 
a local time-varying frequency, 
following the spectral frequency peak. 
It plays the role 
of a variable frequency 
in an amplitude-phase representation 
(the analytic signal) 
which locally best fits the signal, 
and can be also related 
to the mean frequency 
(see \cite{Boa:92}, 
\cite{Co:95}, 
\cite{Boa:15}). 

However, 
the whole spectrum of a signal 
cannot be represented in general 
by a single number, 
so the instantaneous frequency 
is just appropriate for monocomponent signals 
(single oscillations). 
Consequently, 
multicomponent signals need to be decomposed. 
Many algorithms have been proposed for that. 
One of the most successful 
is the Empirical Mode Decomposition, 
which is an algorithm 
for decomposing a signal 
into a finite set of oscillatory signals, 
called Intrinsic Mode Functions (IMF), 
which are 
(argued and supposed to be) 
monocomponent 
and thus suitable to possess 
a meaningful instantaneous frequency 
(via the Hilbert Transform 
and the analytic signal procedure). 
This is the so called 
Hilbert-Huang transform, 
and the resulting 
time-frequency-amplitude/energy distribution 
is called the Hilbert Spectrum 
(see \cite{HSLWSZYTL:98},  
\cite{HS:14}, 
\cite{LBBM:12}). 

The Hilbert spectrum seems to settle 
a more appropriate notion of frequency 
than previous methodologies. 
However, 
in spite of their success, 
these tools have also counterparts. 
The definition and physical interpretation 
of the notions of 
instantaneous frequency, 
as well as the amplitude-phase decomposition, 
the monocomponent signal, 
etc. 
are far from being completely clarified, 
and doubts persist. 
In addition, 
the instantaneous frequency 
at each IMF is not homogeneous 
(it does not need to remain 
into a narrow band), 
so apparently different vibrating modes 
could seem to coexist at each IMF. 
In other words, 
depending on the context, 
the resulting IMF 
are not assured to be meaningful. 
Furthermore, 
the mathematical foundation of 
all these data-driven techniques 
turn to be very challenging, 
and most of the confidence they generate 
comes from numerical experiments 
and particular data analyses. 
There are also technical problems 
to overcome 
(\cite{HS:14}). 


\begin{center}
\textbf{Conclusion}
\end{center}

Frequency is a fundamental concept in Science. 
As it is thought that 
the celebrated inventor Nikola Tesla once said, 
\textless\textless
\emph{if you want to find the secrets of the universe, 
think in terms of energy, frequency and vibration}
\textgreater\textgreater. 

However, 
it is not an easy task 
to rigorously ``define'' 
what frequency means, 
or even should mean. 
Some attempts have firmly been stablished 
in the context of Fourier Analysis, 
with great success and diffusion. 
However, 
some problems arise with these definitions 
in the Time-Frequency Analysis, 
as the uncertainty principles, 
which seem to be counterintuitive 
from a naive perspective 
(see the musical score metaphor 
in \cite{dB:67}, \cite{Gro:01}), 
or quite natural, 
from the perspective of Quantum Mechanics. 
Other attempts, 
as the instantaneous frequency, 
are quite promising, 
but need a more rigorous ground. 

In this regard, 
it is interesting to recover 
the discussion in \cite{Sl:76} 
(see also \cite{Sl:83}) 
about the mathematical modelling 
of ``real world'' signals. 
Slepian distinguishes two constitutive components 
in quantitative physical sciences: 
Facet A, 
for the real world of observations and measurements, 
and Facet B, 
for the abstract world of mathematical modelling 
of Facet A. 
According to Slepian (\cite{Sl:76}), 
\textless\textless
\emph{as usually used [...], 
the words ``bandlimited,'' 
``start,'' ``stop,'' 
and even ``frequency'' 
describe secondary constructs from Facet B of our field. 
They are abstractions we have introduced 
into our paper and pencil game 
for our convenience 
in working with the model. 
They require precise specification 
of the signals in the model 
at times in the infinitely remote past 
and in the infinitely distant future. 
These notions have no meaningful counterpart 
in Facet A}
\textgreater\textgreater. 
Moreover, 
in \cite{Sl:83} he also writes that 
\textless\textless
\emph{it is senseless to ask 
if real signals are bandlimited, 
or timelimited. 
Verification requires real measurements 
at arbitrarily high frequencies 
or at arbitrarily remote or future times, 
experiments that can never be carried out. 
The notions of bandlimitedness 
or timelimitedness belong to 
the engineer's model, 
not the real world 
[...]. 
As it suits him, 
he can assume in his model 
either that his signals 
are timelimited, 
or that they are bandlimited, 
or neither. 
But he should take care 
that the deductions he makes 
from his model about the real world 
do not depend sensitively 
on which assumption he has made}
\textgreater\textgreater. 

This is definitely an old topic, 
but still in vogue. 
In our humble opinion, 
there is a real need 
for generating new strategies 
which could be able 
to fill the previously explained gaps 
between theory and practice, 
between real problems 
and mathematical modelling answers. 
Consequently, 
there are many developments 
to come in this area, 
which will produce 
more appropriate and fruitful notions of frequency, 
susceptible to be applied 
in practice in many disciplines, 
as EEG Analysis. 



\begin{thebibliography}{99}

\bibitem{ABF:94} 
Auscher, Pascal, J. J. Benedetto, and M. W. Frazier. 
Remarks on the local Fourier bases. 
Wavelets: mathematics and applications (1994): 203--218. 

\bibitem{Bab:20} 
Babiloni, Claudio, et al. 
International Federation of Clinical Neurophysiology 
(IFCN)--EEG research workgroup: 
Recommendations on frequency and topographic analysis 
of resting state EEG rhythms. 
Part 1: Applications in clinical research studies. 
Clinical Neurophysiology 131.1 (2020): 285--307. 

\bibitem{Boa:92} 
Boashash, Boualem. 
Estimating and interpreting 
the instantaneous frequency of a signal. 
I. Fundamentals. 
Proceedings of the IEEE 80.4 (1992): 520--538. 

\bibitem{Boa:15} 
Boashash, Boualem. 
Time-frequency signal analysis and processing: 
a comprehensive reference. 
Academic Press, 2015. 

\bibitem{dB:67} 
De Bruijn, N. G. 
Uncertainty principles in Fourier analysis. 
Inequalities 2.1 (1967): 57--71. 

\bibitem{Ch:03} 
Christensen, Ole. 
An introduction to frames and Riesz bases. 
Vol. 7. Boston: Birkh\"auser, 2003. 

\bibitem{CAMcS:06} 
Clifford, Gari D., Francisco Azuaje, and Patrick McSharry. 
Advanced methods and tools for ECG data analysis. 
Boston: Artech house, 2006. 

\bibitem{Co:95} 
Cohen, Leon. 
Time-frequency analysis. 
Vol. 778. Englewood Cliffs, NJ: Prentice Hall PTR, 1995. 

\bibitem{CM:91} 
Coifman, Ronald R., and Yves Meyer. 
Remarques sur l'analyse de Fourier \`a fen\^etre. 
Comptes rendus de l'Acad\'emie des sciences. 
S\'erie 1, Math\'ematique 312.3 (1991): 259--261. 

\bibitem{Coo:87} 
Cooley, James W. 
The re-discovery of the fast Fourier transform algorithm. 
Microchimica Acta 93.1 (1987): 33--45. 

\bibitem{CLW:67} 
Cooley, James W., Peter A. W. Lewis, and Peter D. Welch. 
Historical notes on the fast Fourier transform. 
Proceedings of the IEEE 55.10 (1967): 1675--1677. 

\bibitem{CT:65} 
Cooley, James W., and John W. Tukey. 
An algorithm for the machine calculation 
of complex Fourier series. 
Mathematics of computation 19.90 (1965): 297--301. 

\bibitem{Da:92} 
Daubechies, Ingrid. 
Ten lectures on wavelets. 
Society for industrial and applied mathematics, 1992. 

\bibitem{DGM:86} 
Daubechies, Ingrid, Alex Grossmann, and Yves Meyer. 
Painless nonorthogonal expansions. 
Journal of Mathematical Physics 27.5 (1986): 1271--1283. 

\bibitem{DJJ:91} 
Daubechies, Ingrid, Stephane Jaffard, and Jean-Lin Journe. 
A simple Wilson orthonormal basis with exponential decay. 
SIAM Journal on Mathematical Analysis 22.2 (1991): 554--573. 

\bibitem{DS:89} 
Donoho, D. L., and Stark, P. B. 
Uncertainty principles and signal recovery. 
SIAM J. Appl. Math., 49(3), 906--931, 1989. 

\bibitem{DMcK:72} 
Dym, H., and McKean, H. P. 
Fourier series and integrals. 
Academic Press, New York, 1972. 

\bibitem{FT:20} 
Nicola, Fabio, and S. Ivan Trapasso. 
A note on the HRT conjecture and 
a new uncertainty principle 
for the short-time Fourier transform. 
Journal of Fourier Analysis and Applications 26.4 
(2020): 1--7. 

\bibitem{Fo:97} 
Folland, Gerald B., and Alladi Sitaram. 
The uncertainty principle: a mathematical survey. 
Journal of Fourier analysis and applications 3.3 
(1997): 207--238. 

\bibitem{Gro:01} 
Gr\"ochenig, Karlheinz. 
Foundations of time-frequency analysis. 
Springer Science \& Business Media, 2001. 

\bibitem{HJB:85} 
Heideman, Michael T., Don H. Johnson, and C. Sidney Burrus. 
Gauss and the history of the fast Fourier transform. 
Archive for history of exact sciences 34.3 (1985): 265--277. 

\bibitem{HW:96} 
Hern\'andez, Eugenio, and Guido Weiss. 
A first course on wavelets. 
CRC press, 1996. 

\bibitem{Hir:13} 
Hirsch, L. J., et al. 
American clinical neurophysiology society's 
standardized critical care EEG terminology: 
2012 version. 
Journal of clinical neurophysiology 30.1 (2013): 1--27. 

\bibitem{HSLWSZYTL:98} 
Huang, Norden E., et al. 
The empirical mode decomposition and the Hilbert spectrum 
for nonlinear and non-stationary time series analysis. 
Proceedings of the Royal Society of London. 
Series A: mathematical, physical and engineering sciences 
454.1971 (1998): 903--995. 

\bibitem{HS:14} 
Huang, Norden Eh., and Samuel S. P. Shen. 
Hilbert-Huang transform and its applications. 
Vol. 16. 2nd Edition. 
World Scientific, 2014. 

\bibitem{Ka:68} 
Katznelson, Y. 
An introduction to harmonic analysis. 
John Wiley \& Sons, New York, 1968. 

\bibitem{LP:61} 
Landau, Henry J., and Henry O. Pollak. 
Prolate spheroidal wave functions, 
Fourier analysis and uncertainty--II. 
Bell System Technical Journal 40.1 (1961): 65--84. 

\bibitem{LP:62} 
Landau, Henry J., and Henry O. Pollak. 
Prolate spheroidal wave functions, 
Fourier analysis and uncertainty--III: 
the dimension of the space of essentially time 
and band‐limited signals. 
Bell System Technical Journal 41.4 (1962): 1295--1336. 

\bibitem{LBBM:12} 
Losonczi, Lajos, et al. 
Hilbert-Huang transform used for EEG signal analysis. 
The international conference 
interdisciplinarity in engineering INTER-ENG. 
Elsevier Limited, 2012. 

\bibitem{NS:06} 
Nunez, Paul L., and Ramesh Srinivasan. 
Electric fields of the brain: the neurophysics of EEG. 
Oxford University Press, USA, 2006. 

\bibitem{ORX:13} 
Osipov, Andrei, Vladimir Rokhlin, and Hong Xiao. 
Prolate spheroidal wave functions of order zero. 
Springer Ser. Appl. Math. Sci 187 (2013). 

\bibitem{SCh:13} 
Sanei, Saeid, and Jonathon A. Chambers. 
EEG signal processing. 
John Wiley \& Sons, 2013. 

\bibitem{SchLdS:12} 
Schomer, Donald L., and Fernando Lopes Da Silva. 
Niedermeyer's electroencephalography: 
basic principles, clinical applications, and related fields. 
Lippincott Williams \& Wilkins, 2012. 

\bibitem{Sl:64} 
Slepian, David. 
Prolate spheroidal wave functions, 
Fourier analysis and uncertainty--IV: 
extensions to many dimensions; 
generalized prolate spheroidal functions. 
Bell System Technical Journal 43.6 (1964): 3009--3057. 

\bibitem{Sl:76} 
Slepian, David. 
On bandwidth. 
Proceedings of the IEEE 64.3 (1976): 292--300. 

\bibitem{Sl:83} 
Slepian, David. 
Some comments on Fourier analysis, uncertainty and modeling. 
SIAM review 25.3 (1983): 379--393. 

\bibitem{SP:61} 
Slepian, David, and Henry O. Pollak. 
Prolate spheroidal wave functions, 
Fourier analysis and uncertainty--I. 
Bell System Technical Journal 40.1 (1961): 43--63. 

\bibitem{Sm:90} 
Smith, K. T. 
The uncertainty principle on groups. 
SIAM J. Appl. Math., 50(3), 876--882, 1990. 

\bibitem{SL:05} 
S\"ornmo, Leif, and Pablo Laguna. 
Bioelectrical signal processing 
in cardiac and neurological applications. 
Vol. 8. Academic Press, 2005. 

\bibitem{To:93} 
Tompkins, Willis J. 
Biomedical digital signal processing. 
Editorial Prentice Hall (1993). 

\bibitem{WS:04} 
Walter, Gilbert G., and Xiaoping Shen. 
Wavelets based on prolate spheroidal wave functions. 
Journal of Fourier Analysis and Applications 
10.1 (2004): 1--26. 

\bibitem{WS:05} 
Walter, G., and T. Soleski. 
A new friendly method of computing 
prolate spheroidal wave functions and wavelets. 
Applied and Computational Harmonic Analysis 
19.3 (2005): 432--443. 

\bibitem{Z:59} 
Zygmund, A. 
Trigonometric series. 
Cambridge Univ. Press, Cambridge, 1959. 

\end{thebibliography}
\end{document}